\documentclass[conference]{IEEEtran}
\IEEEoverridecommandlockouts
\usepackage{cite}
\usepackage{amsmath,amssymb,amsfonts}
\usepackage{algorithmic}
\usepackage{graphicx}
\usepackage{textcomp}

\usepackage{subcaption}

\usepackage{listings}
\usepackage{color}
\usepackage{xcolor}

\definecolor{mygreen}{rgb}{0,0.6,0}
\definecolor{mygray}{rgb}{0.5,0.5,0.5}
\definecolor{mymauve}{rgb}{0.58,0,0.82}

\definecolor{backcolour}{rgb}{0.95,0.95,0.92}

\lstdefinestyle{promptstyle}{
    backgroundcolor=\color{backcolour},
}

\usepackage{array}
\newcolumntype{C}[1]{>{\centering\arraybackslash}p{#1}}

\def\BibTeX{{\rm B\kern-.05em{\sc i\kern-.025em b}\kern-.08em
    T\kern-.1667em\lower.7ex\hbox{E}\kern-.125emX}}
\begin{document}
\title{Qiskit Code Assistant: Training LLMs for generating Quantum Computing Code}


\author{
\IEEEauthorblockN{
Nicolas Dupuis\IEEEauthorrefmark{1}, 
Luca Buratti\IEEEauthorrefmark{2}, 
Sanjay Vishwakarma\IEEEauthorrefmark{3}, 
Aitana Viudes Forrat\IEEEauthorrefmark{3}, \\
David Kremer\IEEEauthorrefmark{3}, 
Ismael Faro\IEEEauthorrefmark{3}, 
Ruchir Puri\IEEEauthorrefmark{1} 
and Juan Cruz-Benito\IEEEauthorrefmark{3}}
\IEEEauthorblockA{\IEEEauthorrefmark{1}IBM Research, Yorktown Heights, NY, USA}
\IEEEauthorblockA{\IEEEauthorrefmark{2}IBM Research, Zurich, Rüschlikon, Switzerland}
\IEEEauthorblockA{\IEEEauthorrefmark{3}IBM Quantum, Yorktown Heights, NY, USA}}

\maketitle

\begin{abstract}
Code Large Language Models (Code LLMs) have emerged as powerful tools, revolutionizing the software development landscape by automating the coding process and reducing time and effort required to build applications. This paper focuses on training Code LLMs to specialize in the field of quantum computing. We begin by discussing the unique needs of quantum computing programming, which differ significantly from classical programming approaches or languages. A Code LLM specializing in quantum computing requires a foundational understanding of quantum computing and quantum information theory. However, the scarcity of available quantum code examples and the rapidly evolving field, which necessitates continuous dataset updates, present significant challenges. Moreover, we discuss our work on training Code LLMs to produce high-quality quantum code using the Qiskit library. This work includes an examination of the various aspects of the LLMs used for training and the specific training conditions, as well as the results obtained with our current models. To evaluate our models, we have developed a custom benchmark, similar to HumanEval, which includes a set of tests specifically designed for the field of quantum computing programming using Qiskit. Our findings indicate that our model outperforms existing state-of-the-art models in quantum computing tasks. We also provide examples of code suggestions, comparing our model to other relevant code LLMs. Finally, we introduce a discussion on the potential benefits of Code LLMs for quantum computing computational scientists, researchers, and practitioners. We also explore various features and future work that could be relevant in this context.
\end{abstract}


\section{Introduction}

We are digitally surrounded by Large Language Models (LLMs) that guide us while writing to serve as assistants for several problems, reduce user writing effort, suggest different options for words/sentences to enhance our style, or fix our grammatical errors. 
The same applies in the context of source code, where LLMs are being used in different stages of the software development cycle: from the generation of code to bug fixing, from generating documentation to migration~\cite{survey2024,copilot}.
Generic LLMs such as GPT-4~\cite{openai2023gpt4}, Claude~\cite{claudetechreport},
or Gemini~\cite{gemini} are very good at coding, but smaller code-LLMs can reach almost the same coding skills while being
easier and cheaper to train, and to infer. 
For example, to cite just a few, StarCoder~\cite{starcoder2}, Code Llama~\cite{codellama}, and DeepSeek Coder~\cite{deepseekcoder} all show impressive performances on various code benchmarks. 


There is a growing interest in the application of Artificial Intelligence (AI) methods to enhance the quantum computing field~\cite{preskill2018quantum, biamonte2017quantum}. In recent years, it has become apparent that the majority of research and development efforts in this area have been focused on devising novel quantum algorithms for artificial intelligence~\cite{bharti2020machine,PERALGARCIA2024100619,dunjko2018machine,emani2021quantum,martonosi2019next,baireuther2018machine,bausch2023learning} or integrating classical AI features with quantum systems to optimize specific domains or processes~\cite{nautrup2019optimizing, emani2021quantum,martonosi2019next,nguyen2021deep,2405.13196}. However, there is a noticeable gap in the application of machine learning and classical intelligence systems and algorithms to augment quantum ecosystems and platforms and empowering quantum computing practitioners, a concern that has gained increasing attention within the quantum community~\cite{van2020quantum,moon2020machine,10.1007/978-3-319-91152-6_32}. 

With the proliferation of Large Language Models (LLMs) for code assistance, and considering the previous comments in the field of quantum computing, an area of interest is to develop specialized LLMs for quantum code generation. Quantum code generation poses unique challenges that make it a more complex task than standard code generation:
\begin{itemize}
    \item It requires a basic knowledge of quantum computing
    \item There is limited amount of data and code examples
    \item The field evolves quickly, and new techniques appear frequently and the relevant libraries are updated accordingly.
\end{itemize}

The same challenges apply to some extent to human coders: quantum computing has a high barrier of entry for developers that are not familiar with the field but want to explore its capabilities. With specialized code assistants for quantum, we aim to make quantum computing more accessible to new adopters, and to make development workflows more efficient for current users.

In this work, we introduce specialized LLMs that can work as code assistants for Qiskit SDK~\cite{Qiskit,2405.08810} users. Qiskit is the lead open-source quantum computing framework~\cite{Fund_2023} that provides a comprehensive set of tools, libraries, and documentation for building quantum algorithms, simulating quantum systems, and working with quantum hardware. 

The paper is organized as follows: Section 2 presents the methods and materials employed to train the LLMs, including details about the dataset, training procedures and foundation models used. Section 3 presents a summary of the results achieved, including a description of our evaluation benchmark, how our model compare to others and some prompt results. Section 4 presents some conclusions and future directions. 

\section{Methods and materials}

We start training on top of a Granite code model~\cite{granite-code}, part of a family of decoder-based models for generative AI code tasks. The Granite series of models shows state-of-the-art performance across open Code-LLMs in a variety of coding tasks. Furthermore, Granite models are among the most open models available as of today ~\cite{bommasaniklyman2024fmti}, providing clear details about data, training and architecture.
Our base model is granite-20b-code which uses gpt\_bigcode~\cite{santacoder} architecture, has $20~$Billion parameters, learned positional encodings, multi-query attention,
and a context length of $8192$ tokens. The tokenizer is identical to StarCoder~\cite{starcoder} and has a vocabulary size of $49152$. The Granite base model was pre-trained on $1.6~$T tokens of code data including 116 programming languages.\footnote{We started this work using an early version of granite code which saw less tokens than the recently published moded\cite{granite-code}.}

To improve the performances of the model at generating high-quality Qiskit code, we extend its pretraining with additional Qiskit data containing python scripts, and Jupyter notebooks. We crawled GitHub using its API, searching for all publicly available repositories  with a permissive open-source license that contain the keyword ``qiskit'' in the name or in the description, keeping only the main branch at the latest commit available, and omitting forks. All the data was collected on April 19, 2024. As common with technical SDKs, Qiskit evolves fast and deprecates features often, so training a model with the latest data available ensure compatibility with the latest releases. After collecting data, we filtered out samples with deprecated code, keeping only samples updated after 2022, and applied exact-match deduplication.    
For the Jupyter notebooks, we followed a similar approach to StarCoder~\cite{starcoder2}, and used sentinel tokens to separate out code and markdown fields.
We also filter out cells containing decoded base64 image data. We did not use the output cells from the notebooks.  

After filtering, the total number of tokens is $88~$M, of which $80~$M were never seen by the base model. We set up training data mixing ratios to ensure diversity and quality. Table~\ref{data-distribution} shows the data and token distribution used. The data includes python scripts, and Jupyter notebooks that contain mix of Qiskit tutorials and code. In the table, we highlight the difference between Qiskit Official (qko) and other non-official Qiskit data (qk). The qko data comes from one of the following GitHub organizations, Qiskit~\cite{qk-gh}, Qiskit-Community~\cite{qk-comm-gh}, or Qiskit-Extensions~\cite{qk-ext-gh}, and is considered of the highest quality, hence the large oversampling factors~($10.3$ and $11.2$). We tested different weights and mixing ratios and found the values in Table~\ref{data-distribution} to work the best on our evaluation benchmark and when sampling from the model.       

For extend-pretraining, the data is packed and we use a special token to separate each samples. The total number of tokens after oversampling is $193~$M and we train for $1400$ steps ($\approx3$~ epochs). We use a global batch size of $64$, and a learning rate warmed-up from $0$ to $1\times10^{-5}$ on $140$ steps then decayed with a cosine schedule.

In order to improve natural language understanding, we further instruct-tune the model. We use the octopack approach as described in~\cite{octopack}, mixing chat data from openassistant ($8$k samples) and commit data from commitpackft ($5$k samples). We also add synthetic data specific to Qiskit: $2.7$k question/answer pairs synthetically generated from tutorial using mixtral instruct model~\cite{mixtral}, and $1$k synthetic 
prompt/code pairs, whose execution accuracy were validated using synthetically-generated unittests. We left-pad all the sequences and use a $2048$ sequence length. We train the model for $3.2$ epochs using a global batch size of $32$, a learning rate of $8\times10^{-6}$ decayed using a cosine schedule and a warm-up of $160$ steps. 

\begin{table}
\centering
\caption{Data distribution and token count used for extend pretraining. Each subset has a weight and is oversampled. Total token count in one training epoch is $193~$M. ``qko'' refers to any sample originating from an official Qiskit GitHub organization.}
\label{data-distribution}
\begin{tabular}{p{0.088\textwidth}*{4}{|C{0.07\textwidth}}}
Dataset & Weight & Epochs & Raw tokens (M) & Eff tokens (M) \\
 \hline
 \hline
qko-code & $0.35$ & $10.3$ & $6.5$ & $67.7$ \\
qk-code & $0.3$ & $1$ & $58$ & $58$ \\
qko-notebook & $0.24$ & $11.2$ & $4.1$ & $46.4$ \\
qk-notebook & $0.11$ & $1$ & $20$ & $20$ \\
\hline
\end{tabular}
\end{table}

\section{Results}

\subsection{Qiskit HumanEval}

For testing our model, we have created a new execution-based benchmark called Qiskit HumanEval (QHE). The benchmark is similar to HumanEval~\cite{chen2021evaluating}, including multiple challenging code problems to solve, all based on the Qiskit library. Among the different problems evaluated through the benchmark we can find management of basic quantum code instructions, writing of particular quantum algorithms, handling of executions in either simulators or quantum devices, basic understanding of quantum information theory-related problems or terminology. QHE currently comprises 101 examples. Similar to HumanEval, the context provided to the model consists in a function definition, followed by a docstring which details the task the model is required to solve. Each example also includes a reference canonical solution, as well as unittests in order to evaluate the correctness of the generated solutions. This benchmark will be released later in 2024.

\begin{figure*}[!t]
\begin{center}
\begin{subfigure}[b]{0.48\textwidth}
\begin{lstlisting}[language=python,columns=fullflexible,]
# Import qiskit ibm runtime service and print which backends are available for
# my account. Get backend "ibm brisbane" and define a function in which I
# can show the properties thermal relaxation time, dephasing time and gate error
# of sx gate for the qubit 1 in that backend.

from qiskit_ibm_runtime import QiskitRuntimeService
service = QiskitRuntimeService()
print(service.backends())
backend = service.backend("ibm_brisbane")

def show_properties(backend):
    properties = backend.properties()
    qubit = 1
    print(f"Qubit {qubit} thermal relaxation time: {properties.qubit_property(qubit, 'T1')}")
    print(f"Qubit {qubit} dephasing time: {properties.qubit_property(qubit, 'T2')}")
    print(f"Qubit {qubit} sx gate error: {properties.gate_error('sx', qubit)}")

show_properties(backend)
\end{lstlisting} 
\vspace{-8mm}
\caption{\textsc{granite-20b-code-qk}}
\label{prompts:qhe}
\end{subfigure}
\quad
\begin{subfigure}[b]{0.48\textwidth}
\begin{lstlisting}[language=python,columns=fullflexible]
# Import qiskit ibm runtime service and print which backends are available for
# my account. Get backend "ibm brisbane" and define a function in which I
# can show the properties thermal relaxation time, dephasing time and gate error
# of sx gate for the qubit 1 in that backend.

from qiskit_ibm_runtime import QiskitRuntimeService
from qiskit.providers.ibmq.runtime import UserMessenger

service = QiskitRuntimeService()
print(service.backends())
backend = service.get_backend("ibmq_qasm_simulator")

def show_properties(backend):
    print(backend.properties().to_dict())
    print(backend.properties().qubits[1].to_dict())
    print(backend.properties().gates[0].to_dict())
show_properties(backend)

\end{lstlisting} 
\vspace{-8mm}
\caption{\textsc{deepseek-coder-33b-base}}
\label{fig:clone_deviant}
\end{subfigure}
\qquad
\begin{subfigure}[b]{0.48\textwidth}
\begin{lstlisting}[language=python,columns=fullflexible,]
from qiskit.circuit.library import LinearFunction
from qiskit.synthesis.linear.linear_matrix_utils
import random_invertible_binary_matrix

def get_random_linear_function(n_qubits, seed):
    """
    Generate a random linear function circuit using the input parameters 
    n_qubits, seed and through using the random_invertible_binary_matrix
    method.
    """
    matrix = random_invertible_binary_matrix(n_qubits, seed)
    linear_function = LinearFunction(matrix)
    
    return linear_function
\end{lstlisting}
\vspace{-8mm}
\caption{\textsc{granite-20b-code-qk}}
\end{subfigure}
\quad
\begin{subfigure}[b]{0.48\textwidth}
\begin{lstlisting}[language=python,columns=fullflexible]
from qiskit.circuit.library import LinearFunction
from qiskit.synthesis.linear.linear_matrix_utils
import random_invertible_binary_matrix

def get_random_linear_function(n_qubits, seed):
    """
    Generate a random linear function circuit using the input parameters
    n_qubits, seed and through using the random_invertible_binary_matrix
    method.
    """
    random_invertible_binary_matrix.seed(seed)
    matrix = random_invertible_binary_matrix.get_random_invertible_binary_matrix(n_qubits)
    return LinearFunction(matrix)
\end{lstlisting} 
\vspace{-8mm}
\caption{\textsc{deepseek-coder-33b-base}}
\end{subfigure}
\caption{Output samples generated with \textsc{granite-20b-code-qk} and \textsc{deepseek-coder-33b-base}. In (a) and (b) the models are prompted with an instruction set as a python comment, while in (c) and (d), the models are prompted with the import statements, a function header and a python docstring.}
\label{fig:output-examples}
\end{center}
\end{figure*}

\subsection{Evaluation Results}
We tested the Qiskit model and other baselines on QHE and HumanEval (HE). To compute the execution accuracy, we used the bigcode harness framework~\cite{bigcode-evaluation-harness} and ran the code generated by the models in a docker environment setup with the latest version of Qiskit SDK (at the time of writing this paper, v1.0.2). Table~\ref{he_qhe_results} presents the results. The pass scores were computed on greedy-decoded model outputs. We compare Granite with 3 base models and one instruct model. Of the three base, \textsc{deepseek-coder-33b-base} has the highest QHE pass score at $39.6$\%. The instruct version of DeepSeek Coder is best at HE but the QHE pass rate is slightly lower. \textsc{starcoder2-15b} is also a strong baseline on QHE with a pass score of $37.62$\%. Our granite base model (\textsc{granite-20b-code}) has a QHE pass score of $20.79$\%, however, after extend training, the pass score reaches $46.53$\%, beating all models.
\begin{table}
\centering
\caption{HumanEval (HE) and Qiskit-HumanEval (QHE) pass@$1$ computed using greedy decoding.}
\begin{tabular}{p{0.23\textwidth}*{2}{|C{0.08\textwidth}}}
Model & HE & QHE \\
 \hline
\textsc{CodeLlama-34b-Python-hf} & $52.43$\% & $26.73$\% \\
\textsc{deepseek-coder-33b-base} & $49.39$\% & $39.6$\% \\
\textsc{deepseek-coder-33b-instruct} & $\mathbf{68.9}$\% & $35.64$\% \\
\textsc{starcoder2-15b} & $45.12$\% & $37.62$\% \\
\textsc{granite-20b-code} & $38.41$\% & $20.79$\% \\
\textsc{granite-20b-code-qk} & $36.58$\% & $\mathbf{46.53}$\%\\
 \hline
\hline
\end{tabular}
\label{he_qhe_results}
\end{table}

\subsection{Prompt results}
\lstset{
basicstyle=\normalsize, 
}
In Fig.~\ref{fig:output-examples}, we present examples of prompt queries sent to granite-qiskit (a), (c) and DeepSeek Coder (b), (d).
\lstset{
keywordstyle=\color{black}, 
}
First comparing (a) and (b), DeepSeek Coder, does not correctly follow the instructions provided. It starts by introducing an unnecessary import with ``from qiskit.providers.ibmq.runtime import UserMessenger'', which is irrelevant and outdated. Furthermore, while it does manage to list available backends, it incorrectly chooses ``ibmq\_qasm\_simulator'' instead of the specified ``ibm\_brisbane'' backend. Also, displaying the backend properties, though somewhat informative, misses the mark by not focusing on the specified properties of thermal relaxation time, dephasing time, and sx gate error for qubit 1. Whilst the output is technically correct in a broader context, it fails to address the prompt specifics.
The Qiskit model correctly responds to the prompt. It accurately selects the ``ibm\_brisbane'' backend and correctly defines the ``show\_properties'' function to focus on the requested qubit and gate properties. When the model proposes code to get qubit properties, it shows a good knowledge about quantum computing terminology as it associates the prompt statement ``thermal relaxation time'' to the term ``T1'' or ``dephasing time'' as ``T2''. 
Now looking at prompts (c) and (d), 
granite-qiskit correctly addresses the prompt specifications, creating a ``get\_random\_linear\_function'' method that accurately employs the input parameters of ``n\_qubits'' and seed. This is achieved through the correct application of generating a random invertible binary matrix and leveraging this matrix to construct a LinearFunction, thereby aligning with the prompt. 
DeepSeek Coder's response, while attempting to address the same task, falls short on several fronts. It uses an unnecessary method call ``random\_invertible\_binary\_matrix.seed(seed)'' that disrupts the standard workflow and makes the code unable to run. Furthermore, the misapplication of the LinearFunction object construction, characterized by an erroneous approach to matrix generation, further underscores a fundamental misinterpretation of the task requirements. 

\section{Discussion}
In terms of the utility of such solutions, we view these tools as a potential catalyst that could accelerate the adoption and utilization of quantum computing, particularly among newcomers and students, much like how LLMs have facilitated the learning of programming in classical languages/paradigms~\cite{finnie2022robots}. For more experienced researchers, computational scientists, and similar user personas, we anticipate that LLMs could improve the coding experience, exploration, and overall happiness~\cite{chen2021evaluating, dakhel2023github,barke2023grounded}, although further evaluation is required to confirm this observation~\cite{vaithilingam2022expectation}.
In relation to the public release of these LLMs, the models trained under the auspices of the Qiskit Code Assistant project will be made available to IBM Quantum users in the upcoming months through a suite of services and extensions that can integrate with existing IDEs or various programming environments. Similarly, the evaluation benchmark, the Qiskit HumanEval, will be released publicly to enable other LLMs to compare and enhance their code suggestions in the field of quantum computing.
Enabling LLMs in a rapidly evolving context like quantum computing presents unique challenges. We anticipate continuous updates to new approaches, algorithms, and libraries. A crucial aspect of this project is the ability to regularly update the models to reflect the most recent code, trends, and best practices, ensuring that they remain valuable and relevant for all user personas. As part of this evolving landscape, we expect the need for features such as code explanation, translation between different libraries or versions, automatic test generation and code repair to arise soon.

\section*{Acknowledgment}
The authors thank the data and model factory who collected the code data and trained the base model, in particular Mayank Mishra, Rameswar Panda, Gaoyuan Zhang, Matthew Stallone, and Hima Patel. We also thank Atin Sood for helping setting up the service, and Xuan Liu for support and management.

\bibliographystyle{IEEEtran}
\bibliography{references}

\end{document}